# Nanocarbon for Energy Material Applications: N$_2$ Reduction Reaction

*G. Centi\* and S. Perathoner*

prof. G. Centi[1] and prof. S. Perathoner,[1]

Dept.s ChiBioFarAm and MIFT, University of Messina and ERIC aisbl, V.le F. Stagno D'Alcontres 31, 98166, Messina, Italy

E-mail: centi@unime.it

**Abstract**

Nanocarbons are an important class of energy materials and one of the relevant applications is for the Nitrogen Reduction Reaction (NRR), e.g. the direct synthesis of NH$_3$ from N$_2$ and H$_2$O via photo- and electro-catalytic approaches. Ammonia is a valuable energy or hydrogen vector. This perspective paper analyses current developments in the field, limiting discussion to nanocarbon-based electrodes. The following aspects are discussed: i) induced charge density differences on C atoms associated to defects/strains, ii) doping with heteroatoms, iii) introduction of isolated metal ions, iv) creation of metal oxide(hydroxide)/nanocarbon boundaries and in-situ dynamics, and v) nanocarbon characteristics to control the interface. Discussion is focused on the performances and mechanistic aspects. It is remarked how the results, notwithstanding the large differences in the proposed nature of the active sites, fall all within a restricted range of performances, far from the necessary targets. A holistic approach is emphasized to make a breakthrough advance in the area.

*SMALL, Invited Concept Paper*

## 1. Introduction

"Nanocarbon" defines the broad area of carbon materials, where the presence of an specific dimensionality at the nano-scale levels plays a crucial role in determining the performances.[1] It includes a large range of different typologies of carbon materials, from those more investigated such as single or multi-walled carbon nanotubes (CNT), carbon nanofibers (CNF), fullerenes and graphene (in the different declinations), to the less common types of other carbon forms having a nanodimension, such as carbon nanohorns, nanocoils, nanodiamond, carbon nanodots, onions and various hybrid forms, graphdiyne, etc. This very wide range of different typologies of nanocarbons is due to the possibility of three different hybridizations for carbon ($sp^3$, $sp^2$ and $sp^1$) giving rise to an unique class of nanomaterials not having the equivalent in other classes of materials.[2] Together with other carbon materials having a less ordered structure, such as active carbon and other amorphous carbon materials including carbon black, hard carbon, carbon foams, graphite, glassy carbon, etc. they form a very large family of materials finding application in many different areas, including the relevant and emerging sector of energy materials.[3] The latter includes several frontier areas of application for energy and chemistry transitions, to develop improved materials for energy storage and conversion, catalysis, and other applications. Among the different sectors where advanced carbon materials play a relevant role, the following can be mentioned: batteries, supercapacitors, fuel cells, electrolyzers and photoelectrocatalytic devices, solar cells, catalysis and photo or electrocatalysis.

Global advanced carbon materials market size is over 6 US$ billion (year 2019), with still a significant further expected growing of CAGR (Compound Annual Growth Rate), over >5%, in the next years.[4] Although most of the applications are for engineered composite

materials, the area of energy materials is among those expected to have the faster market increase. Thus, the area of nanocarbons for energy material applications is an industrial and economically relevant area, with large relevance to enable the energy transition and scientifically interesting for the possibility to tailor in detail their properties at the nanoscale allowing thus to obtain significant advances in terms of understanding the structure-reactivity relationships at the nanoscale level.[5-7]

These nanocarbons can be used either as such (metal-free nanocarbons), although typically doped with heteroatoms to improve the performances, or as a support for metals (single atom metals, clusters or larger metallic particles), metal-oxides or other type of nanoparticles. The nanocarbon, however, often does not act as a simple support for these nanoparticles but instead has more complex functions. It acts as charge carrier or to favor surface migration of species[8] (protons, for example), as shuttle for redox reactions in photocatalytic tandem reactions[9], or plays directly a co-catalytic role.[10] For the latter aspects, the functional groups present on the nanocarbon (or even electron poor or rich carbon atoms[11]) can have a direct co-catalytic role (including in photo- and electro-catalytic reactions[12-15]) and/or act as the sites to enhance or modify the chemisorption of the incoming molecules, and/or helping in keeping anchored the intermediate molecules[16] to favor their further transformation (to obtain, for example, >C1 products in the electrocatalytic conversion of $CO_2$[17,18]).

In addition, the nanocarbon (by specific control of the defect sites) may become a semiconductor acting in synergy with the supported semiconductor particles.[15] Modification of the nanocarbon is crucial in all these situations, in addition to play a role in better anchoring the supported nanoparticles, to avoid sintering for example. However, often this interaction leads to significant changes in the nature of the supported nanoparticles, for example by changing their shape.[19-21] Thus, controlling nanocarbons characteristics, not only their shape or dimensionality, but also through their modifications by different ways (doping with heteroatoms, controlled introduction of defects, surface functionalization and creation of nanohybrids, creation of strained regions), is a crucial aspect to improve both the performances and to understand their behavior in nanocarbon-based energy materials. The term nanocarbon engineering describes the different ways to control the characteristics of the nanocarbon to modify its properties and reactivity.

The role of the nanocarbon dimensionality in the performance of these materials for energy storage and conversion has been well reported, for example in the recent roadmap on carbon materials for energy storage and conversion.[22] The role of doping or modification of nanocarbons with heteroatoms in relation to their use as energy materials has been also well reported, for example Meng et al.,[23] Paul et al.,[24] Hu and Dai,[25] Antonietti and Oschatz[26] and Bayatsarmadi et al.[27]. In their progress report, Endo, Terrones and others[28] have analyzed various aspects related to defect engineering and surface functionalization of nanocarbons, but focusing attention on their use as metal-free catalysis. The role of defects sites in nanomaterials for electrocatalytic uses has been discussed by Jia et al.,[29] but centred on developing defect-based motifs as active sites. Zhang et al.[30] have analyzed recently the role of surface functionalization, doping with heteroatoms and creation of defects to generate and control the nature of the active sites in nanocarbon catalytic materials. Zhu and Mu[31] put emphasis on how the introduction of defects in nanocarbons, in synergy with heteroatoms, allow to create potential active sites in nanocarbons.

While certainly the influence of all these aspects on the nature of the active sites is important, other aspects which also depend on the same parameters play an important role, especially in energy conversion applications. They are related to surface mobility, conductivity, hydrophilic/hydrophobic character and creation of triple phase boundaries, which in turn depend on the nanocarbon engineering aspects.[32, 33] However, the discussion in energy-materials for energy conversion and storage is often focused on single specific aspects, such as the nanostructure and dimensionality of nanocarbons and porosity characteristics.[34-41]

One of the relevant areas of nanocarbon as energy materials is that related to the direct synthesis of energy or hydrogen vectors by using renewable energy sources. In fact, the current limitations in renewable energy sources, such as production by photovoltaics and wind, are their discontinuous and fluctuating production, the limits in large storage of electrical energy, and the inefficient transport at long distance. These limitations could be overcome by producing easy storable and transportable energy or hydrogen vectors, which represent thus the key technologies to enable a world-based renewable energy system substituting that based on fossil fuels. To reduce the cost, however, it is necessary to develop direct technologies for the synthesis of these vectors, either by a photocatalytic approach, or by using photoelectrocatalytic (PEC) devices. Two chemicals having this potential to be easy storable and transportable energy or hydrogen vectors are formic acid (HCOOH) and ammonia ($NH_3$), produced from $CO_2$ and $N_2$, respectively, while $H_2$ for their synthesis is produced on-site from water.

The current state of the art is to have separate stages, for example production of renewable electrical energy, followed by an electrolyzer to produce gaseous $H_2$, and then a catalytic step to convert $CO_2$ or $N_2$ in the presence of $H_2$. Direct synthesis means to integrate in a single step/device these three separate steps, but also to avoid to form gaseous $H_2$, and then to activate it for the catalytic conversion, because the hydrogen-equivalent $H^+/e^-$ are produced and used directly on site. This will allow a significant potential increase in efficiency, being eliminated the overpotential in producing and using gaseous $H_2$. In addition, the system inefficiencies in coupling different steps operating at different conditions (and dynamics) are avoided.

Both formic acid and ammonia can be reversibly decomposed to generate $H_2$ (thus acting as hydrogen vectors) or be used directly in fuel cells, turbines, or other devices. The advantage of ammonia, however, is that the hydrogen content by weight is higher (about 18% with respect to about 4% for formic acid) as well as the heating value is higher, but especially ammonia use generates $N_2$, which can be reintroduced in the atmosphere, rather than $CO_2$, which should be recycled to have an effective clean vector. Ammonia is thus an attractive carbon-free energy or hydrogen vector, but its implementation requires to develop direct synthesis routes by using $N_2$, renewable energy and water (NRR, nitrogen reduction reaction). For this reason, the research interest on NRR is increasing. [42-50]

The role of defect engineering in NRR, although not related to nanocarbons, was recently remarked by Yang et al.[51] The role and importance of the electrocatalytic interface for NRR was remarked by Hu et al.[52] Nanocarbons, in view of what discussed above regarding their great flexibility (i) to tune their properties by defect and feature engineering (combining different modalities allowed from their unique features, such as creation of strained regions, functionalization, defect sites, heteroatoms doping, control of -), and (ii) to develop controlled interface properties on conductive materials, are quite interesting components of NRR electrocatalysts.

Nanocarbons such as g-$C_3N_4$ have been reported as interesting metal-free NRR photocatalysts,[53, 54] but a larger range of possibilities can be expected from nanocarbon photocatalytic active materials, or by use nanocarbons to create heterostructured photocatalysts,[55] interface engineered systems with other semiconductors[56] and hybrid

materials.[57]

Thus, a large range of possibilities of nanocarbons in the NRR area can be indicated. While in part analogous aspects have been discussed for $CO_2$ conversion, for example in relation to the electrocatalytic $CO_2$ reduction reaction ($CO_2$RR)[29] or the photocatalytic reduction of $CO_2$,[58-60] or in relation to water splitting,[61] there is still the need to analyze these aspects in relation to NRR.

The aim of this concept paper is thus to highlight some of the possible strategies in developing advanced electro- and photo-catalysts for NRR based on nanocarbon materials, in order to highlight the many possibilities of using them to develop either novel catalysts or promote further the existing ones for this highly challenging reaction which will play a relevant role for the future of sustainable energy and chemistry, if improved energy materials could be developed.

## 2. Nitrogen Reduction Reaction (NRR)

The NRR reaction:

$$N_2 + 6H^+ + 6e^- \rightarrow 2NH_3 \qquad (1)$$

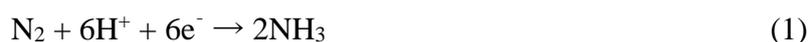

should be coupled with a water oxidation reaction (or other reactions which can equally produce protons/electrons) to close the cycle:

$$3 H_2O \rightarrow 6H^+ + 6e^- + 3/2\ O_2 \qquad (2)$$

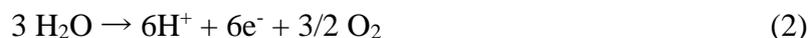

With respect to the catalytic reaction of ammonia synthesis:

$$N_2 + 3H_2 \rightarrow 2NH_3 \qquad (3)$$

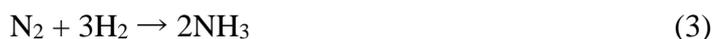

the main difference is that protons/electrons, rather than molecular $H_2$, are used in the ammonia formation mechanism. This means that the overpotentials necessary to both form and use molecular $H_2$ are avoided, with thus a potential higher energy efficiency of the process. There is another main mechanistic difference between photo/electro-catalytic NRR and the catalytic thermal ammonia synthesis (Haber–Bosch process). The latter occurs through a dissociative mechanism. $N_2$ molecules on the catalyst surface (metallic iron) is dissociatively chemisorbed as the first step, i.e. there is a cleavage of the N≡N bond before the occurrence of the addition of hydrogen atoms. Although there are no definitive proofs, the photo/electro-catalytic path instead involves an associative path, with a hydrogen addition to the chemisorbed $N_2$ molecule before the breaking of the N≡N bond. However, the intermediates forming by this addition are high energy molecules, and thus the possibility of an effective path involves the possibility to have a multiple proton/electron simultaneous transfer to the chemisorbed undissociated $N_2$ molecule (forming $N_2H_2$* like species), being lower the energy barrier when simultaneous multiple proton/electron transfer occurs with respect to that of a step-by-step process.[62] This is analogous to the mechanism occurring in nature on *Nitrogenase* cofactor.[62] This is a low energy profile path which can make very effective the transformation, but requires to have special sites able to make possible this multi-proton/electron transfer. In iron-oxide supported on carbon nanotubes electrocatalysts, these multi-proton/electron transfer sites are associated to the presence of an interface boundary between nanocarbons and supported metal-oxide nanoparticles.[62]

There is thus a difference between the mechanisms of $NH_3$ formation from $N_2$ on thermal catalysts (commercial ones are based on metallic iron) and those for electrocatalytic ammonia direct synthesis from $N_2$. This is confirmed by the recent results which has confirmed that oxide/hydroxide iron (γ-$Fe_2O_3$ or FeOOH) species are those active in NRR,[63] rather than metallic (Fe) or iron nitride ($Fe_2N$),[64] which are those identified as the active phases in thermal ammonia synthesis catalysts. The reason is related to the intrinsic difference between thermal operations at high temperature/pressure, which are determined from the need to overcome the blocking effect on the reactivity given by the formation of strong bound N species deriving from $N_2$ dissociation, at photo/electro-catalysis ambient temperature/pressure conditions, where an enzyme-like mechanism dominates based on an electron/proton transfer mechanism on undissociated $N_2$ molecule.

## 2.1. NRR on nanocarbons

Among the many different types of electrocatalysts investigated for NRR, iron-oxide supported on carbon nanotubes (CNTs) show good performances, with best results of ammonia formation rate and Faradaic efficiency of 41.6 $\mu g \cdot mg_{cat}^{-1} \cdot h^{-1}$ and 17%, respectively (after 3h in-situ activation). These are values aligned with the best values reported in literature for NRR[65-70] and thus indicate that these electrocatalysts can be considered representative of the state-of-the-art in this area. There is a maximum in the activity as a function of the loading of iron-oxide on the CNTs.[71,72] However, part of the iron-oxide can be removed by washing without influencing the activity and the part only strongly interacting at defect sites of CNTs, and for this reason stabilized against the easy removal, remains. Removal of other weakly interacting iron-oxide species leads to the enhancement of the specific catalytic behaviour in NRR.[71] This iron-species, initially present in the form of α-$Fe_2O_3$ transforms in-situ to γ-$Fe_2O_3$[71] (from hematite to maghemite). Crystallite nanoparticles have few nm dimensions. Figure 1 reports a high-resolution transmission electron microscopy (HRTEM) image of activated iron-oxide nanoparticles supported on CNTs. Regular fringes are observed in the nanoparticle with a spacing of 0.24 nm, which is the (311) interplanar distance of the cubic maghemite (Fig. 1b). The expansion of the zone of interaction between the iron-oxide particles and the CNT (inset of Fig. 1a) well evidences that the oxide nanoparticles are sitting at carbon defect sites, which stabilize them with respect to sintering and removal from the CNT and induce an elongated shape (thus a modification of the exposed crystalline faces).

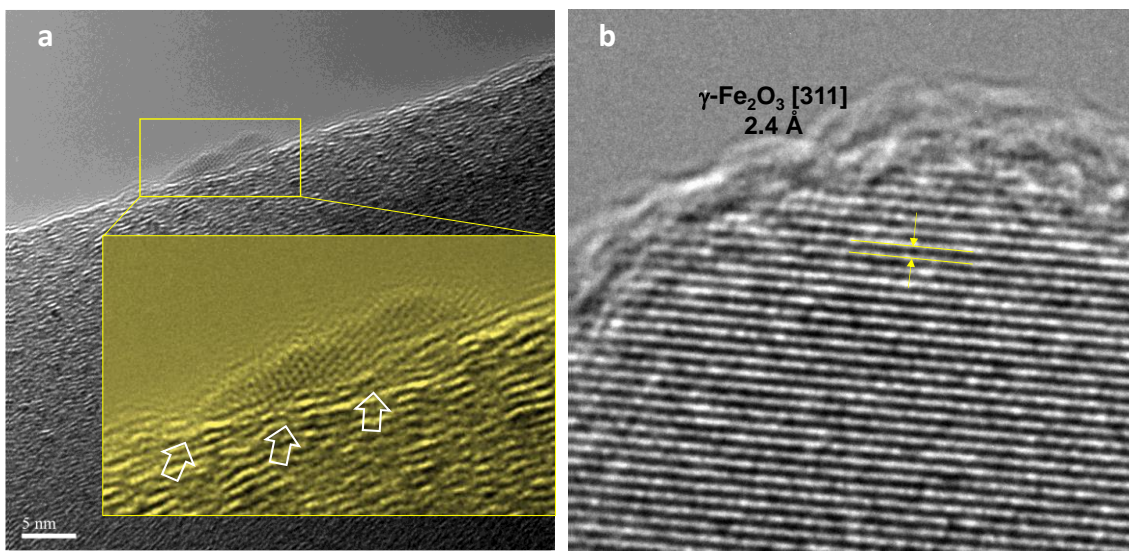

**Figure 1** HRTEM of activated iron-oxide/CNT electrocatalysts for NRR. The iron-oxide nanoparticles are sitting ad defect sites of the carbon nanotubes. The inset of Fig. 1a evidence the presence of many strains in the carbon nanotube graphitic layer due to the interaction with the iron-oxide nanoparticles. Regular fringes are observed in the nanoparticle with a spacing of 0.24 nm, which is the (311) interplanar distance of the cubic maghemite (Fig. 1b). Elaborated from ref.[63] 2020@ Elsevier copyright.

*2.1.1. Role of induced charge density differences on C atoms associated to defects/strains*

The HRTEM image in Figure 1a (inset) shows also that the sitting of the iron-oxide on nanocarbon induces a series of strains in the graphene layers of the carbon nanotube. These strains play likely a role in activating $N_2$ molecule, as shown in the model of $N_2$ activation on carbon strains reported in Figure 2.

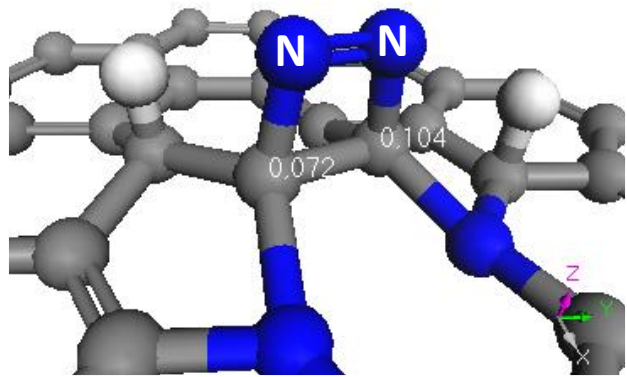

**Figure 2** Model of $N_2$ activation at Stone-Wales defects with nearlying doping N sites (in blue) of a graphene sheet of CNTs. Elaborated from ref.[11] 2018@ Royal Chemical Society copyright.

A common type of nanocarbon defects, particularly in the presence of doping heteroatoms, are Stone-Wales defects. They are constituted by two five-membered rings and two seven-membered rings which substitute four adjacent six-membered rings. The C-C bond bridging two of the adjacent rings results strained with the creation of a reactive C-C bond bridging the two five-membered rings. The presence of nearlying heteroatoms induces a charge density difference at the carbon atoms in this strained C-C bond. In Figure 2, the numbers at the C atoms indicate the effective charge density. The model also illustrates the presence of adjacent protons which could give rise to a concerted $2H^+/2e^-$ transfer on the coordinated $N_2$ molecule, to give a $N_2H_2*$ type species.

The mechanism is similar to that proposed by Chai et al.[73] for oxygen reduction reaction (ORR) at doped Stone–Wales defects on metal-free nanocarbons, with the activity depending on the curvature around the active site with a maximum limiting potential (0.80 V) in the volcano plot for the ORR activity. The reaction mechanism of $O_2$ activation proposed by these authors[73] involves the $O_2$ coordination at these reactive C-C bonds, followed by concerted $2H^+/2e^-$ transfer and formation of a C(OH)-C(OH) species, which then reform the original strained C-C bond by further addition of protons/electrons and water desorption.

There are various findings in literature that support this model. Wu et al.[74] reported the behaviour of O-doped hollow carbon microtubes (metal-free) for electrocatalytic NRR. They observed good performances: Faradaic efficiency of 9.1% at −0.80 V vs. a reversible hydrogen electrode (RHE) and a $NH_3$ formation rate of 25.1 µg·h$^{-1}$·mg$_{cat}^{-1}$ at −0.85 V vs. RHE under ambient conditions. The activity was related to defect sites at the microtube mouths, but where highly strained C-C bonds also exist. On the other side, as outlined in Figure 2, defects and strained C-C bonds, as well as carbon atoms with a defined charge density are strongly interlinked aspects. Even when computational methods have been considered for modelling the reaction sites, it should be verified whether all the possible alternatives have been properly validated together with the level of computational accuracy to exclude alternatives.

Wang et al.[75] have investigated, experimentally and by theoretical studies, NRR on oxygen-doped graphene. They reported a $NH_3$ formation rate of 21.3 µg·h$^{-1}$·mg$_{cat}^{-1}$ and a Faradaic efficiency of 12.6% at −0.55 and –0.45 V vs. RHE. The proposed NRR mechanism was associated to the presence of C=O and O-C=O groups, in combination to defects created which modify the charge density at C atoms acting for the chemisorption of $N_2$ molecules. Data are thus in agreement with the concept presented in Figure 2, also considering that the introduction of heteroatoms and defects in graphene induces deformation from the planarity and thus also strained C-C bonds.

Zhao et al.[76] used oxidized carbon nanotube materials (O-CNT) as electrocatalysts for NRR, observing a $NH_3$ formation rate of 32.3 µg·h$^{-1}$·mg$_{cat}^{-1}$ and Faradaic efficiency of 12.5% at −0.4 V vs. RHE. They indicated that C–O groups play a key role in the electrochemical $N_2$ fixation, but the effective mechanism is also related to the creation of defects and induced charge density on surface C atoms. Huang et al.[77] used oxygen-doped metal-free carbon nanosheets (from carbonization of tannin) as electrocatalyst for NRR. The $NH_3$ formation rate was 20.1 µg·h$^{-1}$·mg$_{cat}^{-1}$ and Faradic efficiency of 5.0 % at −0.6 V vs. RHE. No indications for the mechanism of reaction was given, but results are well aligned to above indications.

It is worth to remark how all these different metal-free nanocarbons, notwithstanding the different source and characteristics, behave quite similarly. The often objection given that the behaviour in metal-free nanocarbons is associated to metal contaminations deriving from the preparation (for example, the catalytic synthesis of CNTs), thus contrast with the evidence that nanocarbon materials deriving from natural or synthesis procedures (very different in terms of modalities and possible presence of metals) result all active.

This similar behaviour, however, suggests a common nature of the active sites in all of them, and the presence of C sites with different charge density, due to strains, defects, insertion of heteroatoms, is the more reasonable interpretation in terms of a common type of active sites responsible for NRR, as well activity in other electrocatalytic reactions, such as ORR.

## 2.1.2. Role of doping with heteroatoms

Figure 2 remarks how doping of nanocarbons with heteroatoms in one of the relevant factors for the modification of the surface properties of the nanocarbon itself causing the creation of reactive sites. However, heteroatoms are not only responsible of the formation of these functional groups (the different possible functional groups are summarized in Figure 3), but their introduction generates also defects, strains and carbon sites with different charge density.

The different functional groups present in nanocarbons by heteroatom doping can be shortly summarized as the following: (i) *Brønsted acido-base sites*, for example carboxylic, phenolic of amine groups, –SO$_3$H sites, –P-OH; (ii) *Lewis acido-base sites*, for example pyridinic N, carbonyl groups, B; (iii) *frustrated Lewis acid–base pairs*, e.g. Lewis acid–base pairs rigidly separated at a short distance without undergoing neutralization;[78,79] (iv) *dangling bonds at the edges*, which are suitable to promote radical chain oxidations, or to activate covalent bonds by interacting with them; and (v) *redox sites*, such as epoxy surface groups able to give redox reactions with incoming molecules.[80] All these sites have been well demonstrated to play a role in many catalytic reactions,[1,11,24,28,81-84] including in electrocatalytic reactions.[85-88]

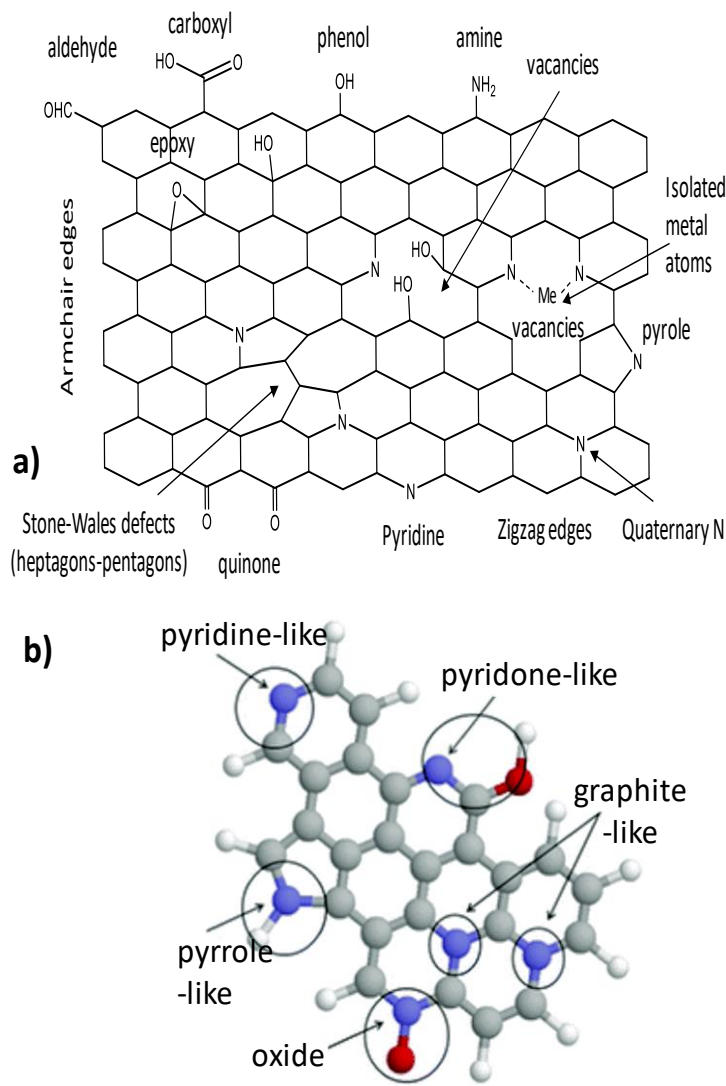

**Figure 3** a) An overview of different possible catalytic active sites in nanocarbons (a graphene sheet as an example). Only doping by O and N is indicated for conciseness. b) Different sites (blue – N, red – O) deriving from N- or O-doping.

*SMALL, Invited Concept Paper*

For a demanding reaction such as NRR, the direct role of these sites, rather than an indirect role of heteroatoms in creating reactive C sites for $N_2$ activation is not enough demonstrated. However, the mechanism of $N_2$ electrocatalytic hydrogenation would require also sites able to transfer electrons and protons to the chemisorbed $N_2$. In metal-free nanocarbons, these functional sites associated to heteroatoms would thus play certainly a role in the overall mechanism of transformation, while the metal-oxide nanoparticles will have likely a role in the electron/proton transfer when they are present (in general, their presence enhances the NRR performance), with thus the active sites being located at the interface between the metal-oxide nanoparticles and the nanocarbon substrate. However, it must be commented that this is a hypothesis, and still not fully proved. Thus, the question of the NRR mechanism, in relation to the effective nature of the active sites, is still open.

Various studies have been dedicated to the effect on NRR of doping nanocarbons with various heteroatoms. Boron-doped graphene have been investigated by Yu et al.[89] They observed best performances for a doping level of 6.2%, with a $NH_3$ production rate of 9.8 µg·h$^{−1}$·cm$^{−2}$ and a Faradic efficiency of 10.8% at −0.5 V versus RHE. The mechanism suggested is that the local electron-deficient environment at the B-doping position offers a strong binding site for $N_2$ adsorption. However, B-doping is also changing the electronic properties of the graphene, creating a p-type semiconductor which increases the electrical resistivity with charge accumulation on B-sites.[90] Thus, when a potential is applied the electron deficient character of B changes and models of the surface reactivity without a proper application of the potential are no longer valid. ORR activity of B-doped graphene decreases with B-doping, e.g. the reduction peak shifts to increasingly negative potentials for increasing level of doping. This is consistent with the charge localization at B sites, and in view of the analogy between ORR and NRR mentioned before, suggests that the idea that B promote

NRR reaction by creating Lewis acid sites (electron-deficient) that coordinate $N_2$ molecules is probably not correct. On the other hand, B-doping also promotes the side, unwanted reaction in NRR, i.e. the transformation of $H^+$ ions to $H_2$ (hydrogen evolution reaction, HER), with respect to undoped graphene.[91] In B-doped graphene, the boron atoms are $sp^2$-hybridized, but the lattice parameters are slightly different causing distortion of the planar structure of the graphene.[92] In addition, boron has one less valence electron than C, which induces a charge polarization in the graphene network, becoming negatively charged. B-doping provides more holes to the valence band of graphene, increasing the carrier concentration and hence improving the electrocatalyst activity, but in the side reaction rather than in NRR.[93] Thus, it is likely that the positive effect of B-doping of graphene on NRR is not related to the substitution of C with some B atoms, but to the creation of other sites.

This is well consistent with the fact that doping with quite different elements in terms of chemical properties, such as P, promotes also the NRR activity. Yuan et al.[95] showed that phosphorus-doped carbon nanotubes represent an efficient metal-free electrocatalyst for NRR with a $NH_3$ formation rate of 24.4 $\mu g \cdot h^{-1} \cdot mg_{cat}^{-1}$, thus higher than that observed for B-doped graphene. The activity was attributed to P atoms substituting C atoms which coordinate $N_2$ molecule, although neither P has the characteristics of substituting C atoms, nor to coordinate $N_2$. Thus, likely the mechanism is related to a different influence of P on the properties of the nanocarbon. Nitrogen and phosphorus co-doped porous carbons were also reported as NRR electrocatalysts,[96] although properties were not excellent with a Faradaic selectivity of 4.2% and an $NH_3$ formation rate of 0.97 $\mu g \cdot h^{-1} \cdot mg_{cat}^{-1}$ and the nature of the active sites was not clearly determined. It is worth to mention that N-doped porous carbons have been also reported as active NRR electrocatalysts, with an ammonia formation rate of 23.8 $\mu g \cdot h^{-1} \cdot mg_{cat}^{-}$

$^1$ at −0.9 V vs RHE.[97] Pyridinic and pyrrolic N were identified as key sites for $N_2$ adsorption and dissociation.

Finally, it could be also reported that Liu et al.[98] have investigated defective carbon-doped boron nitride nanosheets as NRR electrocatalysts. The $NH_3$ formation rate reported was 36.7 µg·h$^{−1}$·mg$_{cat}^{−1}$ at −0.55 V RHE with a Faradic efficiency of 6.5%. The activity was related here to the defective nature of this electrocatalyst. Yin et al.[99] instead indicated by theoretical studies that sp$^2$-hybridized B is the optimal species for providing high activity for NRR, specifically B substituting an edge N atom in the cavity of $C_2N$. Cao et al.[100] proposed instead, also by using a theoretical approach, that single and double B atoms embedded in natural or defect cavities of $C_2N$ are the active sites for NRR. The double boron atom species ($B_2$@$C_2N$) is a more efficient electrocatalyst with extremely low overpotential of 0.19 eV comparing to B@C2N (0.29 eV). It is worth to note that in this structure the N or B active sites discussed above as active centers are also present, but the NRR activity was instead related to different type of centers, which are summarized in Figure 4.

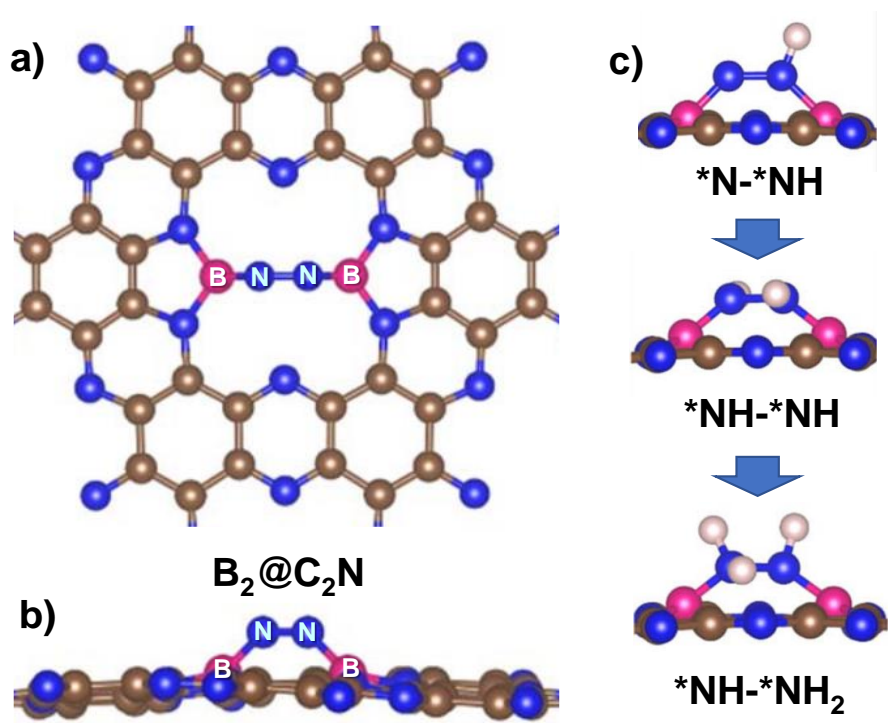

*SMALL, Invited Concept Paper*

**Figure 4** Top (a) and side (b) view of optimized structures of $N_2$ adsorption on $B_2@C_2N$ for end-on configuration. (c) Optimized geometric structures of some of the reaction intermediates in NRR on $B_2@C_2N$. Elaborated from ref.[100] 2019@ IOP Publishing copyright.

S-doped carbon nanospheres[101] have been also reported as active NRR catalysts with a $NH_3$ formation rate of 19.1 $\mu g \cdot h^{-1} \cdot mg_{cat}^{-1}$ and Faradic efficiency of 7.5% at –0.7 V versus RHE. Thus, different types of heteroatoms with quite different characteristics have been reported to promote NRR in different type of nanocarbons, but the overall performances are relatively similar, notwithstanding always they were claimed as superior electrocatalysts. Different roles for the promotion effect have been proposed, although never looking at all the possible situations, and taking into consideration that in spite of the quite different type of proposed mechanisms, the performances are relatively similar.

Note also that few studies have considered the critical question of selectivity,[67] e.g. how to promote NRR, while depress the parallel hydrogen evolution reaction (HER). Zhang et al.[102] proposed that non-metallic catalysts (such a metal-free nanocarbons) show the advantage in electrochemical nitrogen fixation allowing a better design strategy to control these competitive NRR/HER reaction by engineering of surface or interface defects. N, B and S doping are part of this nanocarbon engineering strategy, which require, however, to have a more comprehensive and holistic approach to understand the many effects which are associated to modification of these nanocarbons.

Defect engineering for electrochemical NRR to ammonia remains thus a key tool,[103,104] using unique characteristics in nanocarbon materials, which, however, require better design criteria accounting the many possible factors influencing the behaviour,[51] as will emerge also from the following discussion on reactors configurations and interfaces. Note also that in-situ modifications often have been not accounted with enough details, and it is known, for example, that high-density topological defects could be created in nanocarbons by ammonia treatment leading to an enhanced carbon dioxide electroreduction.[105] It is thus possible that similar defects could be formed also during the electrocatalytic ammonia synthesis.

*2.1.3. Introduction of isolated metal atoms*

$N_2$ adsorption is a relevant element in the NRR mechanism, but also other factors will play a crucial role, such as electron and proton transfer, the latter being discussed in the following part. For electron transfer, a preliminary concept should be commented, which is of general validity for electrocatalytic processes, although often not properly considered. In an ideal perfectly homogeneous electrode, the application of a potential will induce an accumulation of charge on the surface, which will be the driving force for the electron transfer. However, if rather than a fully homogeneous flat surface, we have some points (charge sinks) where charge surface accumulation occurs, they will act as the catalytic points to accelerate the rate of electron transfer. Therefore, nanostructured electrodes will allow to control charge accumulation upon application of a potential (similarly also when charges are photogenerated) and to stabilize these charges for the time necessary to allow the electron transfer to the incoming molecule (the electron transfer to incoming molecules is a slower process with respect to the process itself of charge generation) is an important element in the design of catalytic electrodes. A perfect graphene layer, although conductive, could be not an ideal catalytic electrode, but when we introduce defects and strains, and/or heteroatoms, then we create sites for both adsorption of $N_2$ (as discussed in the previous section) and where charges (electrons) could accumulate and enhance the rate of the electron transfer.

According to this model, the introduction of metal ions in the nanocarbons could thus favor the electron-transfer step in NRR, but on the other hand this is one element in the mechanism that should work in synergy with the other to make effective the reaction. It is thus necessary to consider the problem from this holistic perspective, otherwise results may be difficult to rationalize.

Mukherjee et al.[106] reported that atomically dispersed Ni site electrocatalysts shows an $NH_3$ formation rate of 115 µg·cm$^{-2}$·h$^{-1}$ (loading about one mg$_{cat}$ per cm$^2$) at –0.8 V RHE with Faradaic efficiency (FE) of 21%, but under different conditions of potential and type of electrolyte (about 7% FE at the same conditions of maximum $NH_3$ formation). The electrocatalyst is obtained by pyrolysis of a ZIF-8 precursors, a bimetallic metal organic framework (MOF) of Ni and Zn. Mukherjee et al.[106] reported that this method allows to obtain atomically dispersed Ni sites on the carbon framework with nitrogen vacancies, although characterization data are not fully convincing that effectively only single Ni atoms are obtained. The mechanism proposed, based on DFT, is that the $N_2$ molecule adsorbs on the top of the central Ni atom in an end-on configuration followed by two hydrogenation steps to form *NNH and *NNH$_2$ before the break the N≡N bond and release $NH_3$ (Figure 5). Note that the second hydrogen is added to the same N atom, differently from other proposed mechanisms (see also later).

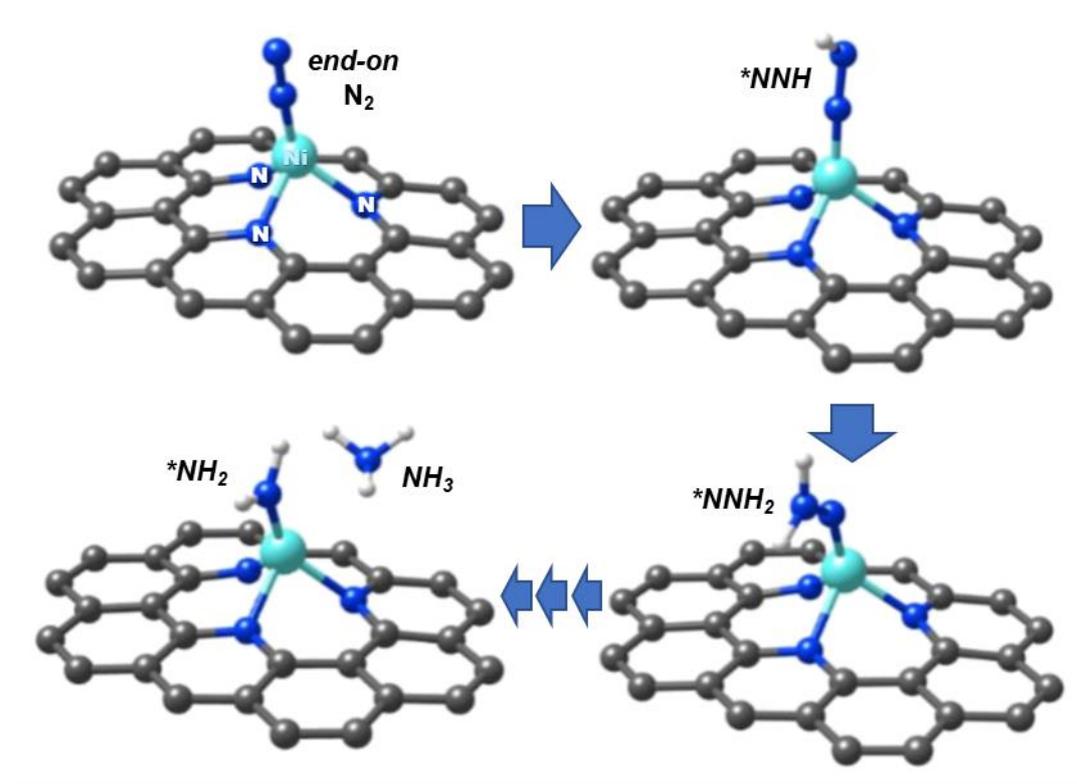

**Figure 5** Proposed pathway of NRR on a Ni-$N_3$ active site derived from pyrolysis of ZIF-8 precursor. In the figure, the gray, blue, white and cyan balls represent C, N, H and Ni atoms, respectively. Elaborated from ref.[106] (Supplementary Info). 2020@ Wiley-VCH copyright.

Besides to the presence of only this type of sites in a material produced by pyrolysis which typically produce a large range of different situations, other unclear aspects are how the proton and electron transfer effectively occurs. Zr is also present in the original MOF and its possible role is also not discussed, neither the nature of the sites giving HER which is the dominant reaction. Low resolution transmission electron images provided do not allow to understand better the effective nature of the environment surrounding the Ni sites, neither whether they are effectively single sites. Thus, several aspects must be further investigated to proof the proposed mechanism as well as alternative possibilities. May be thus still reasonable that the role of Ni is instead to act as sites for favoring the electron transfer necessary to convert chemisorbed $N_2$ in a concerted mechanism synergistically with the proton transfer.

*SMALL, Invited Concept Paper*

A general comment, common also to other studies, regards if the DFT approach is effective in describing correctly electrocatalytic reactions. The predicted energy evolution of NRR on a Ni-N3 site (derived from pyrolysis of ZIF-8 precursor) under an applied potential of 0, –0.3, and –0.8 V versus RHE are shown in Figure 6.[106]

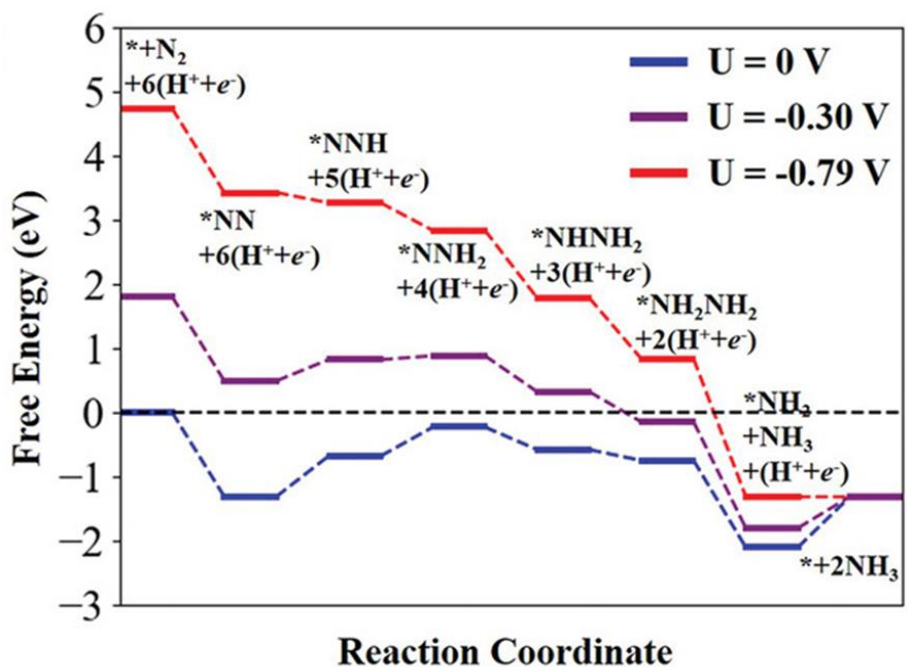

**Figure 6** Predicted energy evolution of NRR on a Ni-N3 site (derived from pyrolysis of a ZIF-8 precursor) under an applied potential of 0, -0.3, and -0.79 V versus RHE. Elaborated from ref.[106] 2020@ Wiley-VCH copyright.

Note that -0.8V is the potential for the maximum productivity in ammonia formation. The predicted energy potential profile indicates a fully thermodynamic feasible reaction without the presence of activation energy barriers and thus proceeding spontaneously, differently from the experimental observations (an increase in the rate on increasing temperature). The authors remarked that the electrolyte may significantly impact the results and estimated profiles, which point out whether the DFT method is accurate in predicting the effective reaction mechanism.

Wang et al.[107] remark instead the need to have heteronuclear dual-atom catalytic elements in nanocarbons, proposing specifically FeMo/g-$C_3N_4$ as an effective electrocatalyst for NRR, being among the 28 evaluated heteronuclear catalysts that have a low limiting potential for NRR (−0.23 V) and a high potential for HER (−0.79 V), thus allowing the highest potential to increase Faradaic selectivity. The optimized FeMo/graphene structure and some of proposed NRR intermediates are presented in Figure 7.

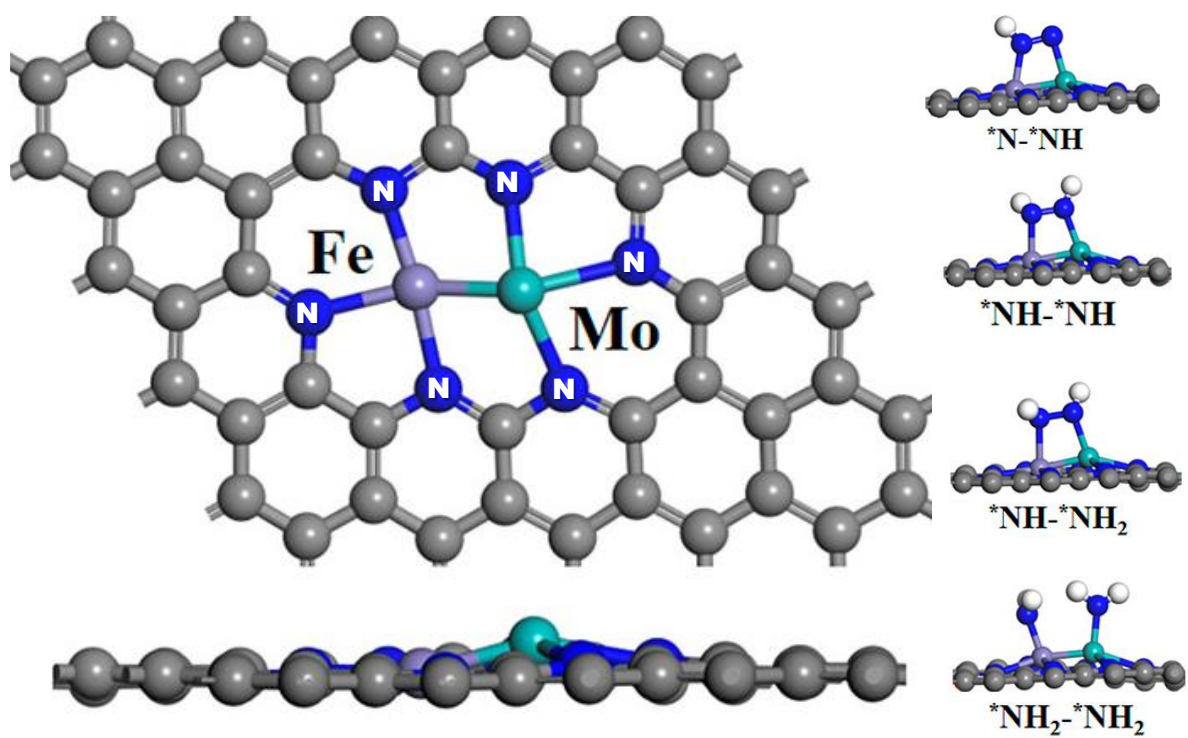

**Figure 7** Optimized FeMo/graphene structure. At the right side some of proposed NRR intermediates. The C, N, H, Mo, and Fe atoms are labeled as gray, blue, white, cyan, and lavender balls, respectively. Elaborated from ref.[107] 2020@ ACS copyright.

Note that with respect to what proposed by Mukherjee et al.[106] here it is indicated a side-on mechanism of $N_2$ chemisorption, and that the second hydrogen transfer occurs on the 2$^{nd}$ coordinated N atoms rather than on the same N atom which is not coordinated with the metal atom. Ni was also one of the metal atoms analyzed by these authors, but they conclude (also

*SMALL, Invited Concept Paper*

by an theoretical approach) that first single atoms tend to aggregate differently from heteronuclear metal dimers and that a dimer is necessary, because the empty d-orbitals of the dual-metal sites can pull two lone pair electrons at both ends of $N_2$, rather than just one lone pair electron of one side of the $N_2$ molecule as for single-site metal ions. However, the effective possibility to have a structure such as that indicated in Figure 7 with sixth N atoms in a graphene defect site, to allow the formation of the heteronuclear metal dimers is questionable.

The metal sites proposed for NRR resemble those present in metal phthalocyanines. Xu et al.[108] studied in NRR an iron- phthalocyanine (FePc) grafted on O-MWCNT (oxygen-functionalized multiwalled carbon nanotubes). This electrocatalyst shows a $NH_3$ formation rate of 36 $\mu g \cdot h^{-1} \cdot mg_{cat}^{-1}$ and a Faradaic efficiency of 9.7%. The proposed mechanism, by DFT calculations, is that the activation of $N_2$ to produce $N_2H^*$ is the rate-limiting step with a ΔG value of 1.79 eV. $N_2$ coordination occurs on the Fe atom.

*2.1.4. Role of the creation of metal oxide(hydroxide)/nanocarbon boundaries and in-situ dynamics*

The third necessary functionality for a NRR mechanism is related to the sites which enable the proton transfer, in a concerted (eventually multi-proton) mechanism with electron transfer. This functionality should be nearby to the sites for $N_2$ (and other intermediates) chemisorption. The DTF proposed mechanisms, as early commented, have not analysed this crucial question, neither in general the need to have a multiproton transfer rather than a step wise addition as in all mechanisms commented above. This is somewhat implicit in the approaches used. To also remark the need of a concerted electron/proton transfer. For example, if the electron transfer occurs before the proton transfer, a negatively charged species forms which facilitates the reaction with protons, but destabilize the intermediate, because the surface is negatively charged for the effect of the potential applied. In the DFT method these aspects are not accounted, but in a real electrocatalyst they are present. If an electrolyte is present, as in most of the tests reported in literature, an electrical double layer is also present, which modifies the surface charge of the electrode, as well as limits the access of gaseous $N_2$ to the electrode surface and also of the protons. For this reason, a great influence of the electrolyte on the performances of NRR is observed, although this is an aspect typically phenomenological and not rationalized.

Note, however, that the liquid electrolyte is not a necessary requirement for NRR. By using an electrocatalytic reactor configuration such as that reported in Figure 8,[71] the electrocatalyst is not in contact with an electrolyte, but only with the $N_2$ gas phase which reacts on the electrocatalysts by diffusing through a gas diffusion layer (GDL) similar to that present in PEM-type (Proton Exchange Membrane) fuel cells. The presence of an additional GDL layer between the electrocatalysts and the membrane is necessary to avoid crossover of the formed ammonia through the membrane[72] and thus presence of ammonia also on the anodic part (where water electrocatalytic oxidation occurs).

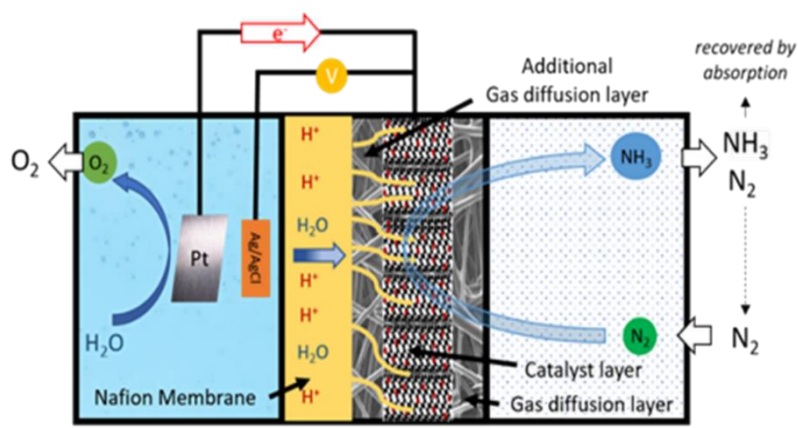

**Figure 8**. Schematic view of the reactor for electrochemical ammonia synthesis with gas-phase configuration for the ammonia synthesis hemicell part. Elaborated from ref.[72] 2017@ ACS copyright.

The great advantages of this electrocatalytic reactor configurations are that i) $N_2$ access to the electrocatalytic surface is not limited by diffusion and solubility in the electrolyte, and ii) the formed ammonia could be easily recovered from the outlet stream of the continuous flow electrocatalytic reactor, differently from other cases where the recovery of ammonia from the electrolyte is typically an energy-intensive operation. Note also that from a practical perspective, this type of cell, avoiding the presence of the liquid electrolyte (as in fuel cells, it can be eliminated also from the anodic part), show great advantages for scalability.

Under the reaction conditions present in Figure 8, the protons, generated on the anodic side diffuse through the membrane and inner GDL to reach the electrocatalysts. Surface diffusion of the protons is a controlling mechanism, differently from the case of protic electrolytes. The presence of sites near those for $N_2$ adsorption able to favour this proton transfer simultaneously to electron transfer is thus a key to enhance the NRR performances. For this reason, iron-oxide on CNTs promotes the NRR reaction. Note that other type of oxide/nanocarbon materials have been reported for this reaction. For example

- $V_2O_3$/C P-doped,[109] showing a $NH_3$ formation rate of 22.4 µg·h$^{-1}$·mg$_{cat}^{-1}$ at −0.35 V vs. RHE, and a FE of 13.8% at −0.25 V vs. RHE
- oxygen vacancy-rich $NiCo_2O_4$ on hollow N-carbon polyhedra[110] showing an $NH_3$ formation rate of 17.8 µg·h$^{-1}$·mg$^{-1}$ with Faradaic efficiency of 5.3%
- vanadium carbide nanoparticles/carbon sphere[111] showing a $NH_3$ formation rate of 34.6 µg·h$^{-1}$·mg$_{cat}^{-1}$ and Faradaic efficiency of 12.2% at −0.40 V vs. RHE
- copper supported on activated carbon functionalized with sulfonate groups (Cu/AC-S),[112] showing a $NH_3$ formation rate of 9.7 µg·h$^{-1}$·mg$^{-1}$ and a Faradaic efficiency of 15.9% at −0.3 V vs. RHE.

Different catalyst families have been also reported, for example $Mo_2C$ nanorods,[113] niobium pentoxide nanowire arrays,[114] and MXene ($Ti_3C_2T_x$, where T = F, OH) nanosheets or nanoribbons,[115-117] among many others. N-free commercial carbon cloth, treated at 450°C in air to create abundant defects, has been also reported as active NRR electrocatalyst.[119] None of them has been reported to show really breakthrough performances, and even if every time the electrocatalysts were indicated as superior to the others, all results were in a quite limited range of variations, as further commented later. Thus, focusing discussion here on iron oxides on nanocarbon materials is not to claim their superior performances with respect to other systems, but only to illustrate some relevant mechanistic aspects.

In fact, an in-situ reconstruction of the metal oxide species may occur during the catalytic reaction. Small (≤ 2 nm) iron-oxide nanoparticles, sitting at defects of N-doped CNTs, transform reversibly during the electrocatalytic reaction because of the application of the potential in combination with the reaction medium.[120] NEXAFS (surface sensitive Fe L near edge X-ray absorption fine structure) operando results combined with ambient pressure XPS data show the in-situ formation of a FeOOH type of species stabilized at carbon defect sites during the electrocatalytic reaction, even if for the $CO_2$ electrocatalytic conversion. Figure 9 reports the proposed optimized *Fh*-FeOOH/N-C interface nanostructure which forms by application of a negative potential within the range of relevance for the electrocatalytic behavior.

Note that the presence of doping N-atoms is relevant to form this species during the electrocatalytic reaction, and, for example, oxygen rather than nitrogen doping of the graphene surface layer of CNT leads to the stabilization of a different surface structure with different (worsen) electrocatalytic properties. In NRR, $Fe_2O_3$/CNT catalysts transform also in-situ, and after removal of the iron-oxide species weakly-interacting with the carbon support, small (≤ 2 nm) iron-oxide nanoparticles, sitting at defects of CNTs, are detected with a maghemite (α-Fe$_2$O$_3$) structure different from the initial hematite (γ-Fe$_2$O$_3$) structure.[63] This transformation leads to about five times increases of both ammonia formation rate and Faradaic selectivity, likely for the in-situ formation of a γ-FeOOH rather than α-FeOOH nanostructure.

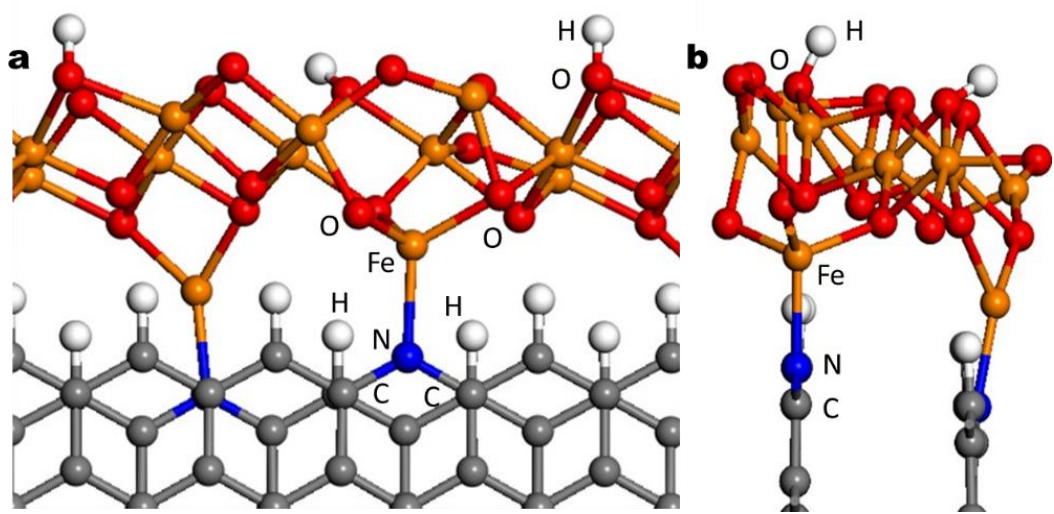

**Figure 9** Model of the *Fh*-FeOOH/N-C interface: a) Top and b) lateral views of the optimized ferrihydrite nanostructures decorating the N-doped graphitic zigzag edges. Color code: C = gray, H = white, N = blue, Fe = orange, O = red. Elaborated from ref.[120] 2018@ Springer Nature copyright.

The relevance of the specific nanostructure derives from the observation, as remarked in Figure 10, of strong analogies between the surface arrangement in γ-FeOOH[62] and the suggested mechanism of reductive conversion of N$_2$ in the *Nitrogenase* FeMo-cofactor.[121,122] In the latter, the presence of a concerted multi/electron and /proton reductive transfer mechanisms is well proved,[121,122] suggesting that γ-FeOOH[62] centers (stabilized and carbon nanotube defects) are able to give an analogous NRR mechanism.

Note that in other electrocatalytic reactions, in particular OER[123,124] that show strong mechanistic analogies with NRR as previously remarked, FeOOH species were also indicated as the active ones. In both cases, the behavior was associated to the specific surface configuration, able to give multielectron transfer.

These iron-oxide/hydroxide nanoparticles are stabilized at carbon defects. HRTEM images reveal the presence of strained C-C bonds and thus likely carbon atoms with charge density at the interface. It is possible that the effective active sites are located at the interface between the nanocarbon and the iron-oxide/hydroxide nanoparticles, with the role of the latter to act as centers to favors the multi/electron and /proton transfer to the chemisorbed $N_2$.

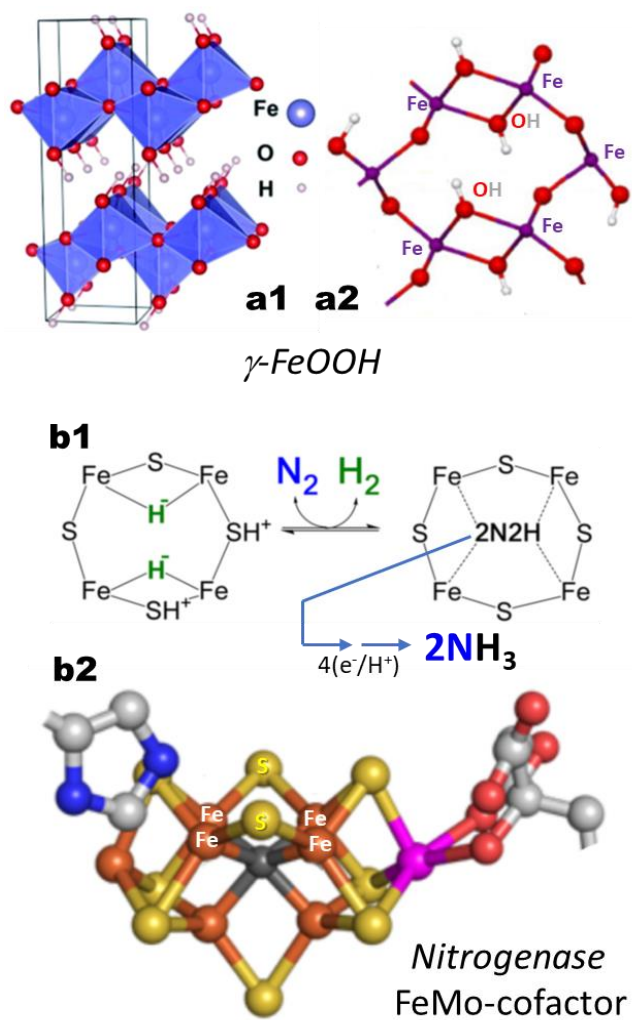

**Figure 10** Proposed γ-FeOOH (a1) nanostructure and related surface structure (a2) for simultaneous multi/electron and /proton transfer in NRR, based on the analogy with (b1) the mechanism of the reductive addition to $N_2$ (to form a diazene-type intermediate, precursor of further conversion to $NH_3$) indicated for the *Nitrogenase* FeMo-cofactor (b2). Elaborated from ref.[62] 2019@ Elsevier copyright.

It should be commented, however, that a careful electron microscopy study of these samples (iron-oxide on N-doped nanocarbons), by combining Bright Field and HAADF STEM methods (probe corrected Bright Field and high-angle annular dark field scanning transmission electron microscopy - HAADF STEM)[120] evidences the co-existence of different type of iron-species: nanoparticles, polyatomic species and single atoms. As commented previously for the Ni case, it is unlikely that only a single type of species is present, and thus to attribute the behaviour to a single type of species present.

*2.1.5. Role of nanocarbon characteristics to control the interface*

A further relevant element to account for a proper analysis of the relationships between nanostructure and surface properties of the electrocatalysts and NRR behaviour is related to the interface characteristics. Hu et al.[52] have recently discussed these aspects in a perspective paper. Three-phase interfaces are a critical issue in electrocatalysis in all processes involving either a gaseous reactant (NRR, $CO_2$RR) or the formation of a gaseous product (HER, OER). A three-phase interface is thus present and thus the electrode interface-architecture construction plays a crucial role for practical applications. For HER and OER, the practice is that gas capping, i.e. the fact that generated gas bubbles remain sticked to the electrocatalytic surface, decreasing the effective available surface of the electrode under operations. For NRR and $CO_2$RR, the three-phase boundary is important particularly for reactor configurations as that shown in Figure 8. However, it is of more general importance, because the optimization of triple-phase interfaces of gas-involving electrocatalysts has three main impacts: exposure of active sites, promotion of mass diffusion and acceleration of electron transfer.[125] Optimizing the interfacial region can allow to control the competitive catalytic reactions.[126] Among the many parameters affecting the catalyst–electrolyte interface may be cited the electrolyte ions, electrode structures, as well as the three-phase boundary at the catalyst–electrolyte interface. The interfacial region modulates the electrocatalytic properties via electronic modification, intermediate stabilization, proton delivery regulation, catalyst structure modification, reactant concentration control, and mass transport regulation.[126] By controlling the interface characteristics by covering the electrode with a hydrophobic, nanoporous polyethylene membrane, Li et al.[127] showed that it is possible to modify the three-phase catalytic interface obtaining a large improvement in the performances, particularly Faradaic efficiency and current density in $CO_2$ electroreduction to CO. By controlling catalyst wettability by using a PTFE modified electrode, Chen et al.[128] changed the amount of $CO_2$ bubbles present at the electrode interface, enhancing the electrocatalytic performances. A fluorine-modified copper electrode, to control interfacial characteristics and wettability was reported to largely enhance current density and Faradaic selectivity in $CO_2$ electrocatalytic reduction.

These examples, among others, well evidence the role of controlling interfacial properties in the electrocatalytic reduction of $CO_2$ and suggest that an analogous impact could be also in NRR. In agreement, changing the availability of protons at the electrocatalyst surface it was demonstrated successful to enhance the Faradaic selectivity in NRR. Suryanto et al.[129] used β-FeOOH nanorods (then reduced into a core–shell α-Fe@Fe$_3$O$_4$ by thermal annealing in a H$_2$ atmosphere) grown over a carbon fiber paper to create a high-surface nanostructured electrode used in conjunction with an aprotic fluorinated solvent–ionic liquid mixture (1H,1H,5H-octafluoropentyl-1,1,2,2-tetrafluoroethyl ether - FPEE) for NRR. The best results were a NH$_3$ formation rate of 0.4 µg·cm$^{-2}$·s$^{-1}$ and a FE of 32% at -0.6 V vs NHE (normal hydrogen electrode), which are very interesting, but with performances strongly depending on the moisture content (in the ppm range) which strongly limits the applicability. Lee et al.[130] increased the Faradic efficiency by 10 percentage points (in NRR with respect to HER) by coating a superhydrophobic metal-organic framework (MOF) layer (specifically, as zeolitic imidazolate framework-71 - ZIF) over the NRR electrocatalyst (Ag-Au).

The electrokinetic control of competitive NRR and HER was the focus of the cited perspective paper of Hu et al.[52] They concluded proposing as novel solution an electrocatalytic reactor largely equivalent to that earlier presented by Chen et al.[71,72] and summarized in Figure 8. As mentioned before, this design allows to overcome the limitations in $N_2$ solubility in the electrolyte, and thus to enhance the $N_2$ adsorption on the electrode surface. On the other hand, this design introduces limitations on the diffusion of protons to the electrocatalytic active sites, and thus it is necessary to design the electrocatalyst to allow a fast surface transport of protons, minimizing at the same time the presence of interfaces which could favours proton recombination with electrons. Thus, interface characteristics different from those necessary when the electrolyte is present are necessary.

## 2.2. Photocatalytic NRR

The photocatalytic direct NRR[53, 131-136] is the alternative approach to the electrocatalytic one combined with a source of renewable energy, for example a photovoltaic cell. The choice of the two options is a matter of debate in general for electrocatalytic approaches, valid also for NRR. In principle, integrating the light harvesting and photocurrent generation in a single device (a photoelectrocatalytic - PEC - unit) or even better in a single catalytic nanoparticle (the photocatalytic approach, which may be visualized as realizing at the nanoscale, typically on different nanoparticle faces, the PEC unit) is the preferable option. Especially the latter will avoid the many interfaces and resistances present when connecting physically separated components. Thus, efficiency is higher in principle. However, have a physical separation between the anodic section (where water oxidation occurs, for example) and the cathodic section (where NRR occurs, or other electrocatalytic reactions, such as $CO_2$RR and HER) has the great advantage that the product of oxidation ($O_2$) is separated from the product of reduction (NH$_3$, for example). This is essential from the practical perspective, for safety reasons and for the difficulty and energy costs to separate downstream the products of reaction.

Thus, physical separation (a PEC type device) is a requirement, but the functional element for the generation of the photocurrent which drives the electrocatalytic transformation can be integrated in the device or can be a separated unit. The anode, in fact, can act both as light harvesting and photogeneration system as well as element for the anodic reaction (for example, by introducing active catalysts for OER). In the other case, a PV cell is used to produce the photocurrent and the potential necessary for the electrocatalytic reaction, with the anode only optimized for the anodic reaction (OER). However, the current density at the potential necessary to drive the full electrocatalytic NRR process (typically > 1.5 V) is at least one order of magnitude higher than that currently possible for semiconductors (below 5-10 mA·cm$^{-2}$). There is thus an intrinsic limit of productivity in the PEC-NRR approach and the current preferable choice is a PV-NRR approach. Directly integrate the PV unit in the NRR electrocatalytic cell, on the other hand, has current limits in terms of stability of the PV material. There is thus the need to develop new innovative solutions to overcome these current limits.

From the mechanistic perspective, the work of Wang et al.[136] may be cited to show that the doping with B of graphitic carbon nitride (g-C$_3$N$_4$) nanosheets could promote the photocatalytic NRR in two different ways, rather than acting directly as active center. B stabilize the photocatalyst exposed active nitrogen atoms (e.g., edge or amino N atoms) against hydrogenation to NH$_3$ (thus is lattice N rather than gaseous N$_2$ which forms NH$_3$) and also suppresses the recombination of photogenerated carriers in g-C$_3$N$_4$. Chen et al.[137] instead indicated that in BCN (an intermediate material between graphite and boron nitride) the edge carbon atoms near to B-N pairs are the active sites toward the NRR, with NH$_3$ formation rate of 7.75 µg·h$^{-1}$·mg$_{cat}^{-1}$) and Faradaic efficiency of 13.8% at −0.3 V versus RHE. However, the direct involvement of lattice N in NH$_3$ formation was not sufficiently excluded and this is a typical problem present in all photo/electro-catalysts containing structural nitrogen.

## 3. Conclusions and Perspectives

A very fast growing of activities on NRR in recent years may be remarked, some of them reviewed here (limited to when nanocarbon materials were used as electrodes). Note that the scope of this perspective review was not to provide a systematic discussion of the topic, but rather to highlight selected aspects to evidence open questions and perspective. However, the concepts presented are of general validity for the entire sector even if the discussion was focused on nanocarbon-based NRR electrodes. Nanocarbons for the higher range of possible tuning of their properties and defect engineering, offer a preferential viewpoint on the topic.

The common approach in most of the publications in this area is to remark the superior performances of their electrocatalysts indicating the nature of the active sites, and reaction mechanism, typically based on DTF or similar theoretical modelling. As shown and commented above, a large variety of active sites and relative role in the NRR mechanism has been proposed in spite of the relatively limited differences in the performances.

This concept is exemplified in Figure 11 which report the best performances for the different reported NRR electrodes (only those discussed in this perspective paper) with respect to the targets indicated for applicability of NRR reaction. The latter are taken from the report "Research needs: towards sustainable production of fuels and chemicals" prepared by the ENERGY-X initiative, now part of SUNERGY initiative.[138] The values of NH$_3$ formation rate (µg·h$^{-1}$·mg$_{cat}^{-1}$) were reported as specific current density of ammonia formation (J$_{NH3}$, mA·cm$^{-2}$) which is a preferable value to monitor electrocatalytic industrial performances. Note that some of the samples discussed here were not reported, because the ammonia formation rate was not possible to be converted to the same unitary value (for example, being missing the electrocatalyst loading on the electrode). Note also that the *x* axis ($J_{NH3}$) is logarithmic.

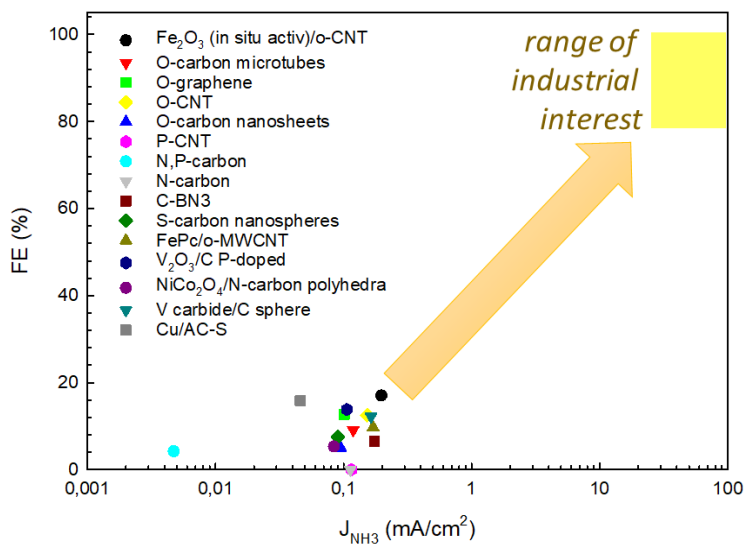

**Figure 11** Specific current density of ammonia formation ($J_{NH3}$) versus Faradaic Efficiency (FE) for the series of nanocarbon-based NRR electrocatalysts reported here, with an indicative range of industrial interest (see text).

Figure 11 clearly evidences that a breakthrough in the NRR performances is necessary to move from the current values to those of industrial interest. However, the Figure also put a question mark on the proposed reaction mechanism and nature of the active sites. As commented above, looking from the perspective presented in Figure 11, the data (and similarly also for other type of electrocatalysts reported in literature) fall within a limited range of values, notwithstanding the great difference in the type of active sites proposed, reaction mechanisms and reaction conditions as well. A different approach would be likely necessary to make a step-forward, and thus also the mechanistic studies likely do not catch the crucial aspects controlling the behaviour, at least in terms of capability to use the indications to design the next generation NRR electrocatalysts. Note that this is not a criticism on mechanistic studies, which are a valuable tool, but to the fact it is still missing the capability to use these indications to progress in the direction of industrial development. It is also necessary to reconsider from a broader perspective the criteria of electrocatalysts design indicated in the various reviews on NRR (for example, vacancy creation, heteroatom doping, facet engineering, composite construction, single-atoms),[132] because, even if may be agreed on them as principle, they are currently unable to make the necessary breakthrough progress.

What is suggested in this perspective paper is that a more holistic approach is necessary. There are many elements which concur in determining the overall behaviour and different relevant steps to consider, from $N_2$ and proton access to the electrocatalytic site, to $N_2$ chemisorption and activation, to hydrogen and electron transfer, preferably in a concerted and simultaneous multi/proton and /electron transfer. Attention has been often focused on single aspects, rather than on a comprehensive vision. A synergy is necessary between the functional elements which activate $N_2$ and make the transfer of protons/electrons to the chemisorbed $N_2$ and reaction intermediate. Charge transport and accumulation, charge separation and light harvesting (in photo-catalysts), the reconstruction of the active sites during the reaction as a consequence of the applied voltage (or surface potential created by the photoinduced charge-separation), the electrode interface characteristics and the effect of the electrolyte (or its absence), the possibility of controlling the interface by a porous membrane coating, the need of concerted mechanisms, are some of the elements highlighted here in the discussion. It is necessary a more comprehensive (holistic) analysis of all these aspects to make the necessary non-linear (breakthrough-type) improvement in NRR. A different approach is thus likely necessary with respect to those currently in use. The integration of all the necessary components of the reaction mechanism is necessary, but with a closer relation and tight integration with an advanced design of the electrocatalytic cells (which includes conditions of operation). This is the synergy necessary for a new design of the electrocatalysts.

Nanocarbon-based electrocatalysts, from this perspective, offer the great advantage of a larger flexibility in terms of modifications and tuning of all the properties necessary to optimize NRR, as commented above. However, attention in literature has been too much focused on claiming superior performances and proposing different reaction mechanisms and nature of the active sites, but not enough supported from a critical correlation with the catalytic results. Figure 11 offers a privileged viewpoint, based on a more general perspective, which allows to understand better the limits of proposed methodological approaches.

Attention has been mainly focused on NRR electrocatalysts, because as commented above, there are still intrinsic limits in productivity by the photocatalytic approach, even if it is attracting the possibility to integrate all elements. They offer the possibility to use solar energy, air and water (also derived from the air itself) to make directly (one step) ammonia, e.g. to develop artificial leaf devices to produce directly energy and hydrogen vectors ($NH_3$) from the air. This is still a long-term challenge, but which can stimulate more novel innovate approaches with respect to those current in use.

**Acknowledgements**

This work was made in the frame of the ERC Synergy SCOPE (project 810182) and PRIN 2017 MULTI-e project nr. 20179337R7, which are gratefully acknowledged.

Received: ((will be filled in by the editorial staff))
Revised: ((will be filled in by the editorial staff))
Published online: ((will be filled in by the editorial staff))

report/. Accessed 1/11/2020.

**Author biographies and photographs**

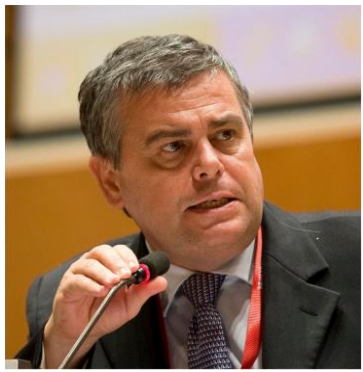

Gabriele CENTI is full professor of Industrial Chemistry at the University of Messina, Italy, President of the European Research Institute of Catalysis (ERIC) and current President of IACS (International Association of Catalysis Societies).

He was coordinator in several EU projects and recently of an ERC Synergy grant on plasma-catalysis. He is board member of SUNERGY. He received several awards, is co-editor in chief of *Journal of Energy Chemistry* and for many years was co-chair of the Editorial Board of *ChemSusChem*. He is author of nearly 500 scientific with current h-index of 83 with over 27.000 citations (Google Scholar).

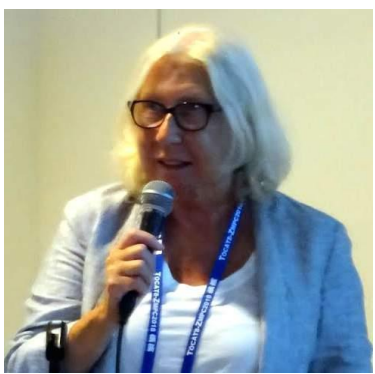

Siglinda PERATHONER gained her PhD in Chemical Science on the photo-physics/chemistry of supramolecular systems with V. Balzani and Nobel Laureate J.M. Lehn. From 2001 she joined the University of Messina where is full professor of Industrial Chemistry. She has coordinated many EU projects and is currently coordinator of the EU project OCEAN on new industrial electrocatalytic paths. Among recognitions the Special Award from "Altran Foundation for Innovation", for artificial leaves for $CO_2$ conversion and the participation to the film "NanoInLife" on nanotechnologie. She is author of about 400 publications. The current h-index is 70 with nearly 19,000 citations (Google Scholar).